\documentclass[conference]{IEEEtran}
\usepackage{graphicx}
\usepackage[T1]{fontenc}
\usepackage{enumitem}
\usepackage{makecell}
\usepackage{subfigure}
\usepackage{xcolor}
\usepackage{color}
\usepackage{booktabs}
\ifCLASSINFOpdf
\else
\fi
\begin{document}

\title{Immediate or Reflective?: Effects of Real-time Feedback on Group Discussions over Videochat}


\author{\IEEEauthorblockN{Samiha Samrose}
\IEEEauthorblockA{University of Rochester\\
ssamrose@cs.rochester.edu}
\and
\IEEEauthorblockN{Reza Rawassizadeh}
\IEEEauthorblockA{Boston University\\
rezar@bu.edu}
\and
\IEEEauthorblockN{Ehsan Hoque}
\IEEEauthorblockA{University of Rochester\\
mehoque@cs.rochester.edu}}

\maketitle

\begin{abstract}
Having a group discussion with the members holding conflicting viewpoints is difficult. It is especially challenging for machine-mediated discussions in which the subtle social cues are hard to notice. We present a fully automated videochat framework that can automatically analyze audio-video data of the participants and provide real-time feedback on \textit{participation, interruption, volume,} and \textit{facial emotion}. In a heated discourse, these features are especially aligned with the undesired characteristics of dominating the conversation without taking turns, interrupting constantly, raising voice, and expressing negative emotion. We conduct a treatment-control user study with 40 participants having 20 sessions in total. We analyze the immediate and the reflective effects of real-time feedback on participants. Our findings show that while real-time feedback can make the ongoing discussion significantly less spontaneous, its effects propagate to successive sessions bringing significantly more expressiveness to the team. Our explorations with instant and propagated impacts of real-time feedback can be useful for developing design strategies for various collaborative environments.  

\end{abstract}

\IEEEpeerreviewmaketitle


\section{Introduction}
Freedom of speech and expression represent the exchange of opinions without fear of retaliation\footnote{https://en.wikipedia.org/wiki/Freedom\_of\_speech}. This cannot be effectively exercised without the parties with opposing viewpoints being respectful to one other during the discussion \cite{Mansbridge2009}. Issues related to politics, religions, race, etc. can infuse heated debate leading to disrespectful or disruptive social behaviors \cite{haidt2012}. If continued, such behaviors can destabilize the discussion and increase the opinion-divide even more. People involved in a heated discussion often misbehave without even fully realizing it \cite{derby2006agile}. Allowing individuals to be aware of their behaviors can impose transparency with the intention of changing the way people react during disagreements. 

\begin{figure}[t]
\centering     
\subfigure[Real-time Feedback Interface. The feedback shown is for the bottom-right participant]{\label{fig:feebackint}\includegraphics[width=0.6\columnwidth]{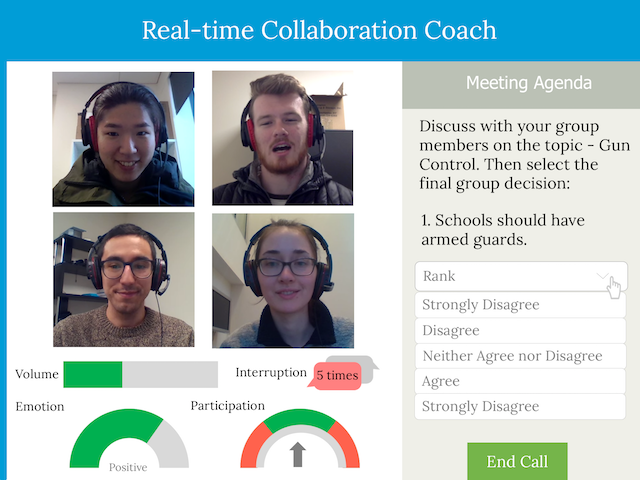}}
\hspace{5mm}
\subfigure[No Feedback Interface]{\label{fig:nofeedbackint}\includegraphics[width=0.6\columnwidth]{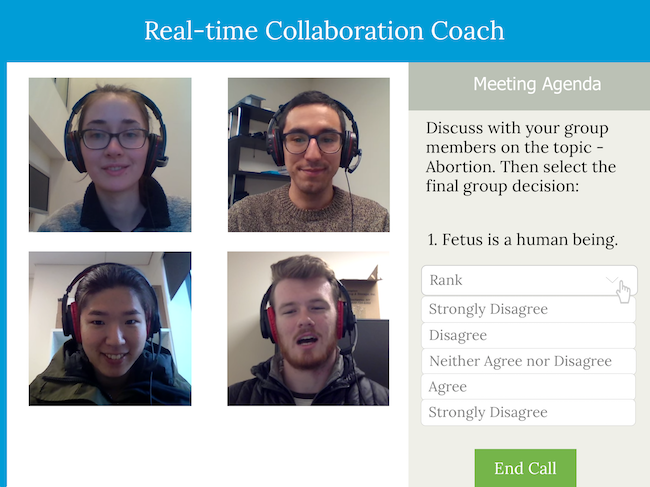}}
\caption{Videoconferencing Systems with Activated/Deactivated Feedback. (a) shows feedback enabled UI with 4 features: volume, interruption, facial emotion, participation or talktime. The features are updated in real-time for each participants.}
\vspace{-0.5cm}
\end{figure}

Video conferencing has become a popular replacement for group conversations to avoid travel, coordination, and requirement for a physical location. The COVID-19 pandemic in 2020 has brought ``stay at home'' orders and travel restrictions, increasing videoconferencing based discussions\footnote{https://www.marketwatch.com/story/zoom-microsoft-cloud-usage-are-rocketing-during-coronavirus-pandemic-new-data-show-2020-03-30}. While the option of videoconferencing may seem more convenient than a face-to-face conversation, it lacks many of the important elements of a face-to-face conversation \cite{sellen1995}. For example, during a video conference, participants are unable to make direct eye-contact, utilize peripheral vision, feel the sense of being co-located and have a hard time inferring and synthesizing the nonverbal cues of other participants. However, the video screen being an integral part of videoconferencing presents an opportunity for showing feedback to the participants, which is not an innate part of the face-to-face conversation setting. Imagine the possibility of promptly and privately reminding the individuals of their subtle behaviors cues that they may not be aware of. How to design such feedback and how participants engage, interpret, and reflect with such automated feedback in short-term as well as long-term open up new research opportunities for the affective computing  community.

Designing real-time feedback for specifically heated debate-like discussions over videoconferencing holds some technical challenges: (1) None of the existing videochat system (e.g., Zoom, Google Hangouts, Skype, etc.) allows real-time analysis of audio-video data, nor do they provide any application programming interface (API) to perform modifications in the client interface to include external feedback; (2) Prior research views \cite{tanveer2015, ofek2013} real-time feedback as distracting which limits the design and the number of feedback features. If not delivered properly, during an ongoing discussion the feedback can overwhelm the users, triggering significant cognitive overload \cite{homer2008}. In this paper, we address these challenges by designing and implementing a working prototype.

We develop a videochat platform integrated with real-time feedback for group discussions. Based on literature review and platform specific design considerations, we select our feedback features: \textit{participation, volume, interruption} and \textit{facial emotion}. Our system extracts the audio-video data to compute and process the feature values for feedback generation. For feedback representation, we design individualized visualization \ref{fig:feebackint} assuring that the performance scores are private to individuals. We evaluate the system in a controlled study with discussion topics on debate infusing issues in the context of USA: \textit{Gun Control, Abortion, LGBTQ, Animal Testing}. Each group participates in two successive discussion sessions having two independent topics. In the first session, the treatment groups receive automated real-time feedback, whereas the control groups just watch a TedTalk video titled "10 Ways to Have a Better Conversation" before starting the discussion (no feedback during the conversation). In the second session, none of the groups receive any feedback. We collect the responses of the participants from two surveys and an interview to further explain the behavior changes. To observe the \textit{instant effect} of the real-time feedback, the performance and the responses of both groups for the first discussion sessions are compared. To analyze the propagated effect of the real-time feedback, we compare both first and second sessions' performances and responses of both groups. We define the second effect as the \textit{reflective effect} of the real-time feedback.

We analyze the \textit{immediate} and the \textit{reflective} effects of the real-time feedback by using both system-captured and self-reported data. The system-captured data shows that while receiving real-time feedback the treatment group participants become less expressive (i.e., less talking) than that of the control group participants with no feedback. The result from the self-reported data shows that real-time feedback makes the treatment group participants more self-conscious. This presents an opportunity for a trade-off between behavior modification and spontaneity of interaction while using real-time feedback. For the immediate reflective effect of the real-time feedback, the system-captured performance analysis shows an increased expressiveness (i.e., more talking) during the second session for the treatment group with no feedback resulting in faster ice-breaking than that of the control group participants. The self-reported data also sheds light on how aware the participants were of their own and group-mates' behavior features. The findings open up new opportunities to utilize the carryover (or priming) effect of real-time feedback for different groups and topics in various settings include in-situ. As numerous research areas with telehealth, job interview, education, etc. can benefit from using our platform to better explore the aforementioned research directions, we have made our code public for other researchers to build on. In summary, the contributions of the work are as follows:

    \begin{itemize}

    \vspace{-0.1cm}
    \item Implementation of a video conferencing system that automatically captures and analyzes a participant's audio and video data. In addition, the interface contains real-time feedback visualizing participation, interruption, volume, and facial emotion metrics.     

    \item Validation using a controlled experiment with 40 participants reveals that while real-time feedback can make the ongoing discussion significantly less spontaneous, its effects carry over to future sessions bringing significantly more expressiveness. 

\end{itemize}


\section{Related Work}
\label{sec:relwork}

\subsection{Behaviors in Heated Discussion} 
For the case of heated group discussions, various key behaviors contribute to escalating or resolving disrespectful or non-collaborative interactions. Identifying these crucial behaviors is the first step towards addressing them. Firstly, mutual respect for the group members is the key to a safe exchange of conflicting ideas \cite{maia2016, Mansbridge2009}. Mansbridge et al. \cite{Mansbridge2009} suggest that participants involved in a difficult conversation should treat one another with mutual respect by actively listening and speaking in a way that helps comprehend the reasoning. Mutual respect can get hampered by dominance and unperceptive behaviors \cite{James1992}. Two key signs of projecting dominance in conversational setting are talking more and interrupting others \cite{dunbar2005, lamb1981nonverbal, burgeon2002nonverbal}. Burgeon and Hoobler \cite{burgeon2002nonverbal} observe that the amount of talking time plays a role in perceptions of dominance and credibility. Dunbar et al. \cite{dunbar2008} show that during conflict infusing conversations people interrupt more to take control and project dominance. Thus, we identify talktime and interruption as two key features to consider for feedback in a heated discussion.

Negative emotion and affect also play important roles in difficult conversations. Jung \cite{Malte2016} shows that affective imbalance during conflict interactions can negatively affect team performance. Anderson and Pearsson \cite{anderson1999} explain the concept and the factors of \textit{incivility spiral} in workplace. They show that repairing negativity is crucial to prevent intense aggressive behaviors from escalating during interactions. This negativity can be expressed prominently through two ways: (a) voice, and (b) face \cite{picard2001, rothenberg1971anger}. We dive deeper into more prior work to address the feature of emotional negativity expressed through these two ways. Research emphasizes vocal tone as a key factor in heated discussions. Rothenberg \cite{rothenberg1971anger} explores the role of verbal hostility as an invariable accompaniments of anger. The research discusses that as a part of expressing anger, the involvements of muscle tension, vascular changes, and involuntary voice change are observed. Negative facial emotion and high vocal tone thus contributes to elevated anger, hostility, and dominance towards other people present in the exchange \cite{sullivan1973clinical}. Derby et al. \cite{derby2006agile} mentions that people, out of anger or excitement, may shout or yell without realizing that they have raised their voices. The study suggests that a gentle feedback can be effective to mitigate the raised vocal volume. Costa et al. \cite{Costa2018} externally manipulated voices to be calmer during conflicts, which resulted in less anxiety among participants. Reflecting on these findings, we identify facial emotion and volume of the vocal tone as another two influential factors for heated conversations.

Overall, research has significantly emphasized features like balanced participation \cite{ab01}, emotional stability \cite{Neubert1998, Nadler2003, derby2006agile}, gesture and language similarities \cite{Kihara2016, david2011}, etc. for coordinated teamwork. Related research done on automated analysis of performance \cite{rawassizadeh2015, ofek2013} and emotion \cite{zeng2009, bailenson2007, keltner2003facial} have paved the way towards tracking group behavior. Generally, maintaining a balance in these features brings better outcome in the form of performance or satisfaction for teams. For example, even though equal speaking time or participation is not always ideal for groups, it generally brings a better experience \cite{Leavitt1951, Diamant2009}. Derby et al. \cite{derby2006agile} discusses how emotional stability can be befinicial in a team setting, which can be achieved by being aware of positive, neutral, and negative emotions altogether. Therefore, providing information about these three zones of emotion can help people having a broader insight to better help maintaining the stability. Burgoon and Newton \cite{Burgoon1991} observe that not actively participating in an interaction bars the feeling of being a part of the ongoing experience. On the other hand, active participants feel more immersed in the interaction and the overall experience \cite{dunbar2008, Burgoon1991}. These suggest that imbalanced participation can affect the group dynamics and overall experience. Therefore, people need to carefully pay attention to both under and over participation to maintain a balance, as both can generate negative impact during conversation in a group \cite{sullivan1973clinical, derby2006agile}. 


Therefore, in the light of prior work, we (1) identify four crucial features for heated discussion: talk-time, interruption, facial emotion, and vocal volume, and (2) recognize that an overall understanding (participation: over, balanced, under; facial emotion: negative, neutral, positive; voice: low, balanced, high) of the behaviors projected by each feature is needed to avoid the negative impact in a heated discussion.

\subsection{Strategies for Real-time Feedback} 

Extensive research has been done with real-time feedback for improving social interactions. Especially, personalized feedback through wearable and cell-phone technologies has been found useful for different personal skill improvement \cite{Boyd2016, hegde2015, rawassizadeh2015,rawassizadeh2018} and emotion analysis \cite{zeng2009, bailenson2007, keltner2003facial}. \textit{MOSOCO} \cite{RN11} provides step-by-step feedback in realtime to handle real-life conversations using mobile phones. \textit{Rhema} \cite{tanveer2015} provides real-time feedback through Google Glass to improve public speaking performances. However, during interactions real-time feedback has been found to be distracting \cite{tanveer2015, ofek2013}. 
Campbell and Kwak \cite{campbell_} find the use of technology while having a conversation with a stranger in public not to be distracting. However they suggest that, to reduce distraction the technology should be carefully designed to be a "natural" part of the face-to-face interaction. Ofek et al. \cite{ofek2013} shows that as the amount of external information increases within the real-time feedback, the distraction becomes more intense. Tanveer et al. \cite{tanveer2015} explores textual and chart-based visualization for real-time feedback for public speaking. Their focus group unanimously agree that showing only a raw performance score without any temporal history is insufficient. Thus, if not designed carefully, the real-time feedback can hamper the discussion negatively instead of assisting the interactions. After receiving real-time feedback, sometimes it can be difficult for users to instantly modify the behaviors \cite{Tausczik}, and sometimes it can be effective \cite{tanveer2015}. But if this feedback is designed properly, it can resolve the problematic interaction properties right when they get tracked. This can decrease further intensification for the case of a heated discussion. Otherwise the problems can intensify to such an extent that it may not be redeemable by pre- or post-feedback. Pre-feedback is useful for scenarios where training is beneficial, whereas post-feedback is useful to evaluate the performance of an already completed experience \cite{grimmett1988reflection, husu2007developing}. These two feedback strategies cannot fulfill the need to immediately address a problematic behavior - the need, as discussed above, is crucial for heated discussions.

Based on all these, we associate two key points from the related literature to justify the possibility of addressing heated discussion using real-time feedback: (1) prior work in Section 2.1 suggests that in a heated discussion showing disrespectful behaviors, addressing the problematic features right when they occur bears the potential to salvage the ongoing conversation. Therefore, the feedback of the design needs to immediately attract the attention of the user. (2) Previous research works discussed in Section 2.2 find that feedback provided before or after the discussion does not address the issues on-spot. However, for the case of real-time feedback, prior work discussed in Section 2.2 emphasizes how real-time feedback attracts user's attention to itself during an ongoing conversation. By combining these two points, we select real-time feedback for our system as it has the potential to be useful in this special discussion with heated conversation.


\subsection{Automated Feedback for Groups} 

For the virtual discussion environment, research explored feedback techniques and influences on virtual groups \cite{Diamant2009, leshed2013, Leshed2009, Nowak2012}. Diamant et. al \cite{Diamant2009} demonstrated that feedback on attention, affective states, etc. resolved potential conflict and dominance in intercultural collaboration setting. Leshed et al. \cite{leshed2007} presented \textit{GroupMeter} showing that in a chat-based discussion environment feedback on linguistic behaviors increased self-focus. Research by Tausczik et. al \cite{Tausczik} regarding suggestive feedback on language usages for chat-based discussions found that, even though the feedback improves group performance, too much negative feedback can hurt the discussion. Nowak et al. \cite{Nowak2012} provided feedback on voice arousal for negotiations conducted over phone. They found that the real-time feedback on one's own vocal modulation negatively affects the user performance.

Even though videoconferencing is a virtual environment, unlike chat or phone-based discussions the interactions exchanged over this platform contain combined audio-visual-textual properties. Although videoconferencing based discussion has similarities with face-to-face discussion, it lacks attention to several key characteristics of face-to-face interactions, such as- rapport, non-verbal cues, etc. \cite{sellen2010, sellen1995}. Addressing heated discussion for videochat setting is crucial for two reasons: (1) The usage of videochat as a discussion medium has increased in recent days among different types of users for different purposes \cite{buhler2013and, ames2010making, judge2010sharing}. (2) For computer-mediated communication, Lea and Spears [40] show disinhibition as one of the central themes and its consequential tendency towards anti-normative behaviour. This is especially crucial for heated debate in computer-mediated communication, since talking more or dominating the conversation is found to occur more frequently in computer-mediated discussions than the face-to-face ones \cite{Losada1990, mcgrath1994groups, straus1996getting}. Therefore, it is important to explore how proper behavior modulation can be facilitated in a heated discussion over videochat platform.

Adopting the same features and feedback styles of other virtual or even face-to-face setups for this platform may not be appropriate. Kim et al. \cite{Kim2008, Kim2012} presented \textit{Meeting Mediator} having a sociometric badge capturing group dynamics showing real-time feedback on participation and turn-taking for videochat based discussions. Notably, the behavior sensing and the feedback showcasing are not integrated parts of the videochat platform. It also deals with very few feedback features to sufficiently explain the group dynamics. Byun et al. \cite{byun2011} designed automated real-time feedback for two-party conversation on a wide range of topics over videoconferencing. They found the positive effects of increased awareness within participants because of real-time feedback, even though the topics are rather diverse. Faucett et al. \cite{Faucett2017} introduced \textit{ReflectLive} providing real-time feedback on non-verbal behaviors in videochat for clinician-patient conversations. He et al. \cite{Helen2017} provided post-discussion feedback on total and acknowledgement words usages, frequencies of smile and eye-contact for videochat-based groups. Samrose et al. \cite{Samrose2018} presented \textit{CoCo: Collaboration Coach} that provided post-feedback on participation, turn-taking, valence, etc. for videochats. \textit{CoCo} includes a good number of features but that may not be applicable for real-time cases. Because unlike post-feedback systems, real-time feedback systems with extensive number of features can impose increased cognitive overload \cite{homer2008, engstrom2017, tillman2017}. Real-time feedback can also differ from post-feedback as the latter has the option to process data after the discussion session, which allows delayed analysis.

The related work so far pointed out three unique gaps that we attempt to fulfill in this paper. Firstly, the videoconferencing is different than face-to-face or chat-based conversations. Thus the real-time feedback for this platform needs to adjust to its needs and norms. We target the information exchange over this medium and explore the effects particularly for such platforms. Secondly, real-time feedback for such a platform can be overwhelming to the users as group dynamics have so many inherent features. Thus we limit the number of features making the information flow limited, resulting in reducing the distraction. Finally, we explore how the real-time feedback affects the users when the feedback in not present afterwards.


\section{Methods}

\subsection{Research Questions}

\vspace{1mm}


\vspace{1mm}
\textbf{\textit{RQ1: What is the instant effect of real-time feedback on groups having heated discussion over videochat?}}

Upon receiving real-time feedback during the ongoing conversation, participants get the opportunity to modify their behaviors on-spot. For a heated discussion over videochat, we want to explore how the participants react to the feedback and what behavior modifications are made. To answer this, we design a videochat system incorporated with real-time feedback, organize a treatment-control study stimulating a heated discussion session, and then compare those sessions. 

\vspace{1mm}
\textbf{\textit{RQ2a: Does the real-time feedback have any propagation effect on future discussions?}}

We want to explore whether effect of real-time feedback in one session propagates to successive sessions even when no feedback is present. The presence of propagation would mean that it works as an indirect feedback, because the participants modify their behaviors by reflecting on the received feedback. We term this effect as the \textit{reflective feedback}. We design a within subject study in which the previous treatment groups have a second discussion session without any feedback. We compare their first and second sessions to capture the presence of any changes. From causality perspective, the changes may also come from other factors (such as icebreaking delay). So we conduct a within subject study with the previous control group in which they again receive no feedback in the second session, and then we measure the changes. The comparison of the final changes of treatment-control groups ensures that the only change left is caused from the \textit{reflective feedback}.

\vspace{1mm}
\textbf{\textit{RQ2b: If RQ2a is "yes", then what is the effect of \textit{reflective feedback} on groups in successive discussions?}}

When evaluating this case, we know from answering RQ2a that there appears a difference in behaviors even when real-time feedback is absent in the successive session. Now we want to observe how the participants react to the \textit{reflective feedback} and what behavior modifications are made. To answer this, we compare the first and the second sessions of the treatment group.



\vspace{-1mm}
\subsection{Feature Definitions}
\label{subsec:designconsideration}

From related literature discussed earlier, we identify highly emphasized four specific feedback features: \textit{participation, interruption, volume,} and \textit{facial emotion}.

\textbf{\textit{(1) Participation:}} The percent amount of time a person talks within a certain time-period. We define it as a combination of (a) talk-time: amount of time spoken, and (b) turn-taking: exchange of speaker's floor. Setting a certain time-period incorporates turn-taking parameter in \textit{participation}. During heated conversation people tend to hold the speaker's floor longer than expected. This feedback feature is to nudge people have a balance in speaking and exchanging turns.  

\textbf{\textit{(2) Interruption:}} The number of times two (or more) people simultaneously talk for a certain time-period. People interrupt others as an impulse from disagreement during a conflict. If for a consecutive time-period two people continue cutting off each other's speech, then both of them are deliberately responsible for the interruption. So we combine both (a) interrupting someone, and (b) getting interrupted in defining the final \textit{interruption}.

\textbf{\textit{(3) Volume:}} The loudness of the vocal tone. In a heated discussion, people tend to subconsciously raise their voices (i.e., increased volume) out of excitement or anger. Shouting at each other escalates the conflict, so we include this feature in our feedback design.

\textbf{\textit{(4) Facial Emotion:}} Compound emotional valence expressed using facial features in the range of positive to negative through neutral. During intense conflict, negative facial emotion such as anger, disgust, etc. can further deteriorate the situation. We incorporate feedback on facial emotion to make people aware of their expressed facial valence.

\vspace{-1mm}
\subsection{\hspace{0.1cm}Feedback Design Considerations}
\label{subs:designconsideration}

As discussed elaborately in Section~\ref{sec:relwork}, we apply real-time feedback in our system to attract the attention of the user towards the problematic behavior on-spot, with a view to observing the behavior changes. We adopt individualized feedback visualization approach for our deign. For every feedback feature, a user sees their own behavior metrics. We adopt the design choice for two reasons: (1) Keeping a user's feedback private to him/her shields from peer-bias. For example, if a group is explicitly reminded that one user is interrupting others a lot, it may influence how the group interacts with the particular user. This can create unintended bias even when that particular user is not interrupting. (2) This empowers the users by letting them decide on modifying their behaviors themselves, instead of enforcing a social pressure on them. For example, if a user knows that everyone in the group is getting notifications about his/her over-participation, then the user may start speaking less because of the social pressure, not because s/he thinks the feedback is appropriate.


\subsection{System Implementation Overview}

\subsubsection{\textbf{\hspace{0.1cm}Videochat System}}
Our videchat system captures and processes the transmitted audio-video data in real-time and generates the performance scores in the backend. Both server and client sides of the backend are involved in real-time data processing to ensure the just-in-time feedback delivery. The Node.js-based\footnote{https://nodejs.org} web socket server connects the client applications ensuring data transmission and message transfer. Hosted in the Microsoft Azure cloud platform, the server uses open-source WebRTC \footnote{https://webrtc.org} for high quality audio-video communication and Traversal Using Relays around NAT(TURN) server \footnote{https://tools.ietf.org/html/rfc8155} for highly secured data exchange. The client application, implemented with JQuery \footnote{https://jquery.com} and HTML 5\footnote{https://www.w3.org/TR/html52}, processes its audio and video feed locally, instead of transferring and processing them in the server. The locality ensures that - 1) each user's performance scores are kept private from other users, 2) the processing overhead is reduced by relieving the server from computing every client's performance scores, and thus faster analysis is achieved. During experiments, we provide the participants with our lab laptops and headphones to maintain the same configurations and settings. The laptop configuration is i5-8265U CPU, 8GB RAM, 256GB SSD, 4K display, 15-inch screen. Each headphone is of high quality with strong brass and high precision in its speaker and hearing units. In the study the client applications run on Google Chrome to maintain the same browser setting.

Even though our videoconferencing system can hold sessions remotely, for our user studies the participants were brought to the lab for providing a homogeneous experimental environment and running different parts of the experiment smoothly. As one of the main purposes of the studies is to verify the sheer presence of the feedback effect, the user study requires maintaining a controlled homogeneous setup. For this reason, we choose the in-lab user study setting.     

\subsubsection{\textbf{\hspace{0.1cm}Feature Analysis}}
\textit{Participation, interruption} and \textit{volume} features are extracted from the audio, whereas the \textit{facial emotion} is extracted using Affdex SDK \cite{McDuff2016} from the video data. We compute \textit{participation} from the percent amount of time a person speaks within the last four-minute window to ensure that they are both speaking and exchanging speaking turns. We pick the four-minute window from testing beforehand for similar fifteen-minute discussions. The feedback of \textit{participation} is shown using three zones: low (under-participation) visualized using the color red, mid (equal participation) using green, high (over-participation) using red. Literature suggests that equal talk-time brings better discussion experience in general, so the mid is the balanced coveted zone. For four users in a discussion, maintaining a static 25\% participation value all the time is too restrictive. So we set a flexible range of 20-30\% for equal talktime range. Thus the low/mid/high zones fall in the ranges of 0-19/20-30/31-100 (\%) respectively. For facial emotion, at each timeframe Affdex SDK provides a score and an associated label for the face (-100 <= negative < 0 = neutral < positive <= 100) which we readjust to (0 <= negative < 50 = neutral < positive <= 100) while visualizing to user. However, frequent negative feedback for a very slightly negative facial emotion can become too discouraging on the users. Thus we re-categorized it as 0-44: negative (red), 45-55: neutral (yellow), 56-100: positive (green). Volume thresholds were computed by converting the extracted microphone volume range in percentage. We exclude noise by removing any value >= 1. The rest of the volume range is divided into three zones for feedback: 1.1<=low(red)<=7, 7.1<=mid(green)<=20, 20.1<=high(red)). These ranges are computed upon testing on multiple users beforehand. For interruption, we set the cutoff time to 1.7 seconds, which means if two users speak at the same time for at least 1.7 seconds it is considered as an interruption for both users. This threshold was set by testing with sentences like "I agree/You are correct" etc. However, during the experiment this threshold was found to be too restrictive and disruptive for the discussion. Therefore we re-set it to three seconds to increase the tolerance. We acknowledge that the assumptions made on conversational heuristics are context dependent and may not generalize. For broad use cases, we envision users to set some of these parameters using mutual consensus or prior heuristics and tweak them as necessary as universal values for some of these metrics would not exist.

\subsubsection{\textbf{\hspace{0.1cm}Feedback Visualization}}

We design two front-end interfaces in the client side of the video conferencing system - (1) With the real-time feedback based on the performance scores, and (2) Without projecting any feedback. Fig~\ref{fig:feebackint} shows the feedback activated interface with four feedback features. The feedback implements graph visualization approach using cascading style sheet (css) and D3.js. For the no-feedback interface shown in Fig~\ref{fig:nofeedbackint}, the analysis is run in the background and the scores are captured, but the feedback visualization is deactivated. This provides us with the relevant metrics, and also makes sure that both interfaces have the same client side processing overhead resulting in the same video conferencing experience except for the feedback aspect.

\subsection{Discussion Topic}
Our discussion topics were designed to infuse a debate-like conversation. The study is conducted in US, so our design topics are targeted to address the current controversial topics there. We picked four main topics, each with three to five sub-topics. The main topics were: (1) Gun Control, (2) Abortion, (3) LGBTQ, (4) Animal Testing. During sign-up, the prospective participants provided 5-scale likert chart responses for each of the subtopics shown in Table~\ref{tab:topic}. Using these responses, we formulate groups to have two discussion sessions in which the group-members hold highly polarized conflicting viewpoints in both discussion topics. Having more topics allows us to better match such groups. For each group, we picked two main topics for two discussion sessions. In each session, the group had to finalize their decisions based on the 5-scale likert chart on three of the given sub-topics. Fig~\ref{fig:feebackint} shows a discussion session, where the main topic is Gun Control with three sub-topics are (two hidden under the dropdown box). The target of the discussion is to decide as a group to choose from a 5-scale likert chart ranging from "Strongly Agree" to "Strongly Disagree" for each sub-topic. It is to be noted that the outcome whether participants converge on a decision is not our target. Rather, the target is to apply feedback so that the group-members maintain balanced respectful collaborative behaviors even when they may never shift their viewpoints on the conflicting topics.


\begin{table}[t]
    \caption {Discussion Topics and Sub-topics} \label{tab:topic} 
    \vspace{-0.6cm}
\begin{center}
  \begin{tabular}{l}

  \toprule
  Topic\\
  \toprule
    
    Gun Control: \\ 
    a. Background checks to allow purchasing guns is an invasion of privacy.\\	 
    b. The Second Amendment protects an individual right to  possess\\ a firearm.	\\
    c. The owner of the gun has the right to bear it everywhere.	\\
    d. More gun laws means fewer gun deaths. \\	
    e. Schools should have armed guards.	\\
    \hline
    Abortion:\\ 
    a. The government can impede a woman's decision to terminate\\ the pregnancy.	\\
    b. A fetus is a human being.	 \\
    c. Doctors have the right to refuse providing information related to\\ abortion procedures.\\
    d. Abortion based on genetic abnormalities is discrimination.	\\
    e. The fetus has the right to live inside the woman's body.	\\
    \hline
    LGBTQ: \\
    a. Marriage is defined as being between a man and a woman.	\\
    b. Legalizing same-sex marriage encourages nontraditional relationships\\ like incest.	\\
    c. Organizations have the right to deny services to homosexual people.	\\
    d. Homosexual couples are incapable of providing their children a\\ stable family.	\\
    e. Sexual orientation is a choice.									\\
    \hline 
    Animal Testing:\\ 
    a. Animal testing should be allowed for treatment purposes.	\\
    b. Drugs that pass animal tests are not necessarily safe for humans.	\\
    c. Animal testing is cruel and inhumane.\\
    \bottomrule
  \end{tabular}
  \vspace{-0.5cm}
\end{center}
\end{table}



\section{User Study}

\vspace{-0.1cm}
\subsection{Participants}

A total of 40 participants were recruited for the study. Each group (i.e., control, treatment) had 20 participants. During the sign-up, the prospective participants provided their stands or polarities on each of the discussion topic/sub-topics. Based on the responses, the discussion groups were optimized to have members with opposite stands on issues. Each group consisted of four members, and each discussion session was 15-minute long. The male and female ratio of the participants was 11:9. All the participants were aged between 18 to 24 years. The ongoing/highest education level was:  11 with some college credits, 24 with Bachelor's degree, 2 with Doctorate degree, 1 with Master's degree.

\vspace{-0.1cm}
\subsection{Study Procedure}

As our objective is to explore the real-time effect of the system, the impact of the visualization property needs to be stabilized across the groups. Therefore, the control group should also be presented with a visual stimulus for receiving the related feature information, except that it would not be in real-time. For this reason, we choose to show a highly influential TEDTalk video to the control group participants before the first discussion session begins. Any video itself is a visual stimulus; on top of that the TEDTalk video also has the speaker showing slides where information was presented using visualization. Thus, by having visual stimuli for both groups, the only observable factor remains to be the real-time nature of stimuli.

\begin{figure}[t]
\centering
\includegraphics[width=0.5\textwidth]{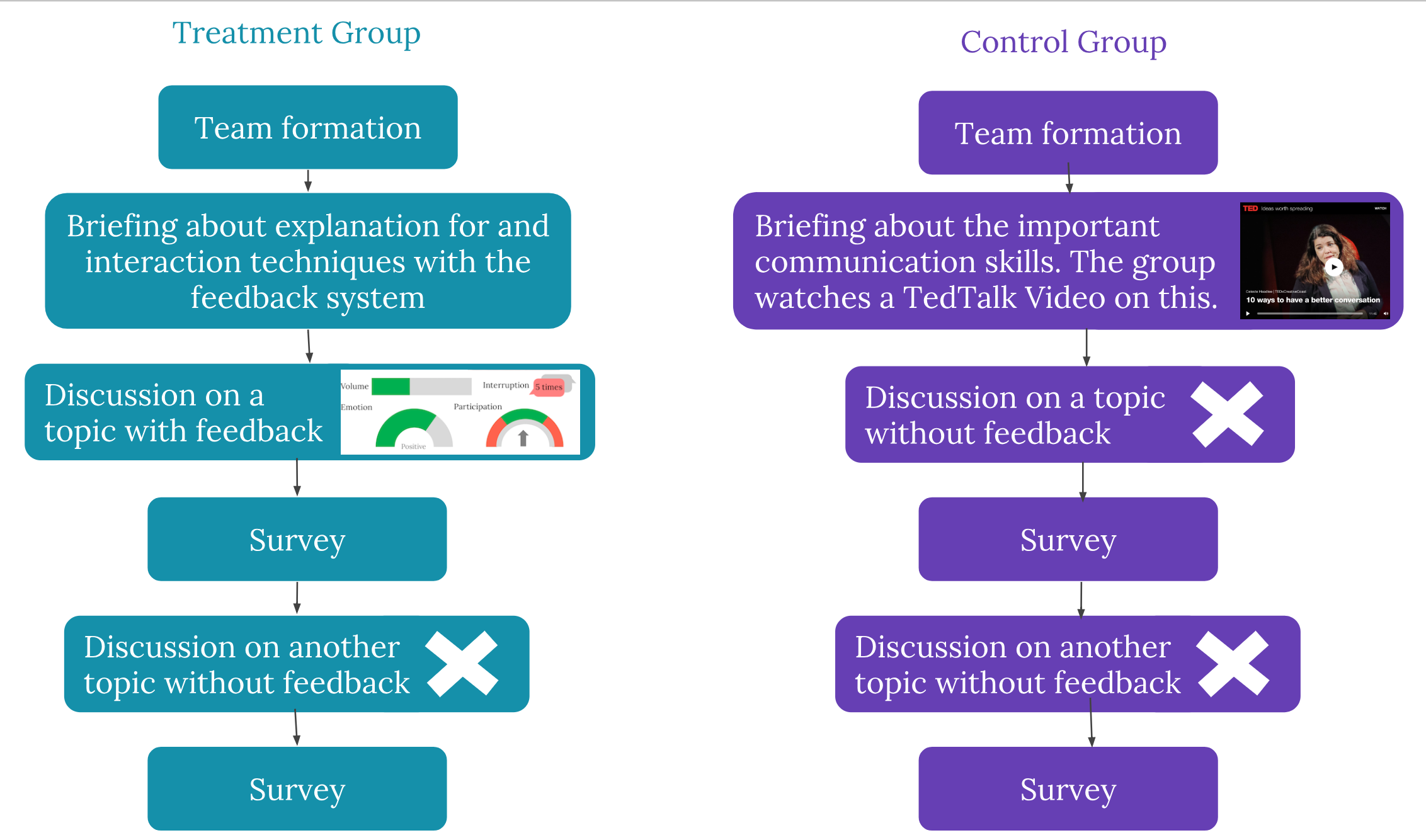}
\caption{Flowchart of the User Study}\label{fig:c-t-outline}
\vspace{-0.5cm}
\end{figure}

Fig~\ref{fig:c-t-outline} shows the outline of our human study. Once we formed the groups based on the sign-up responses, we randomly divided them into control and treatment groups. Both the control and the treatment groups had two discussion sessions on two different discussion topics. Each discussion session was about 15-minute long. Before session-1, the treatment group received a briefing on the feedback system and its features. On the contrary, the control group watched a TEDTalk Video titled "10 Ways to Have a Better Conversation"\footnote{https://www.youtube.com/watch?v=R1vskiVDwl4} by Celeste Headlee. This popular video, which has 22M views on TED.com, was selected as it discusses the relevant features present in our designed feedback system. The reason behind showing the video to the control group was to stabilize the primary feedback knowledge for both treatment and control groups. After the debriefing, four participants of a group went to four different rooms to have the videochat discussion. None other than the participant was present in each room. In session-1, both groups participated in a videochat on a provided topic to reach a unanimous decision. The difference is that the treatment group received real-time feedback integrated in the video conferencing system, whereas the control group received no such feedback from the system. Upon finishing the session, each participant received an online survey regarding the discussion and the performances. Next, the groups participated in another video conferencing based discussion on a different topic. In this session-2, neither the treatment nor the control group received any real-time feedback. We designed session-2 with no-feedback to observe the after-effect of the real-time feedback and compare the treatment group's effect with the baseline control group. After completing session-2 discussion, the groups filled up another survey. The survey questionnaires are designed based on the previous research materials used by Leshed et al. \cite{Leshed2009} and Samrose et al. \cite{Samrose2018}. In the end, we conducted a semi-structured interview for each group. 


\section{Results}
We examine both (1) the machine captured performance features, and (2) the self-reported data of the participants. This section presents the comparisons and the findings of the corresponding data.

\subsection{System-captured Performance Analysis}

We present the comparisons between the control and treatment groups in their two sessions for each of the performance feature metric. The Low-Mid-High attributes represent the feedback divisions or zones. As per our experimental design, we apply 2 (control/treatment) X 2 (session-1/session-2) two-way ANOVA with $\alpha=0.05$.

\subsubsection{\textbf{\hspace{0.1cm}Result Overview}} Two-way ANOVA conducted to examine the effect of experimental setup condition (control/treatment) and state condition (session-1/session-2) on \textit{low} performance attribute shows that there was a statistically significant interaction effect, $F(1,76)=4.73, p=0.03$. For \textit{mid} and \textit{high} performance, no effects were statistically significant. Applied similarly for \textit{facial emotion} metric, for \textit{negative} and \textit{neutral} there were no differences in any of the effects. However, for \textit{positive} attribute, the control-treatment main effect was statistically significant with $F(1,76)=6.53, p=0.013$. This means that we can further explore the changes in \textit{low performance} and \textit{positive facial emotion} attributes across conditions.

In accordance with our three research questions, we divide the comparisons into three different phenomenons:- (1) RQ1 answered from \lq\lq Control vs treatment groups' session-1\rq\rq: to measure the effect of our designed real-time feedback on the participants' performances in comparison with no-feedback environment; (2) RQ2a from \lq\lq Compare session-1 vs session-2 of the treatment group, and the same for the control group\rq\rq. It is to verify the presence of the after effect of the feedback on the groups; (3) RQ2b from \lq\lq Control vs treatment groups' session-2\rq\rq: to compare and measure the reflective feedback on participants' performances.

\subsubsection{\textbf{\hspace{0.1cm}Evaluation of RQ1}}

\begin{figure}[t]
\centering     
\subfigure[Participation Performances]{\includegraphics[width=0.48\columnwidth]{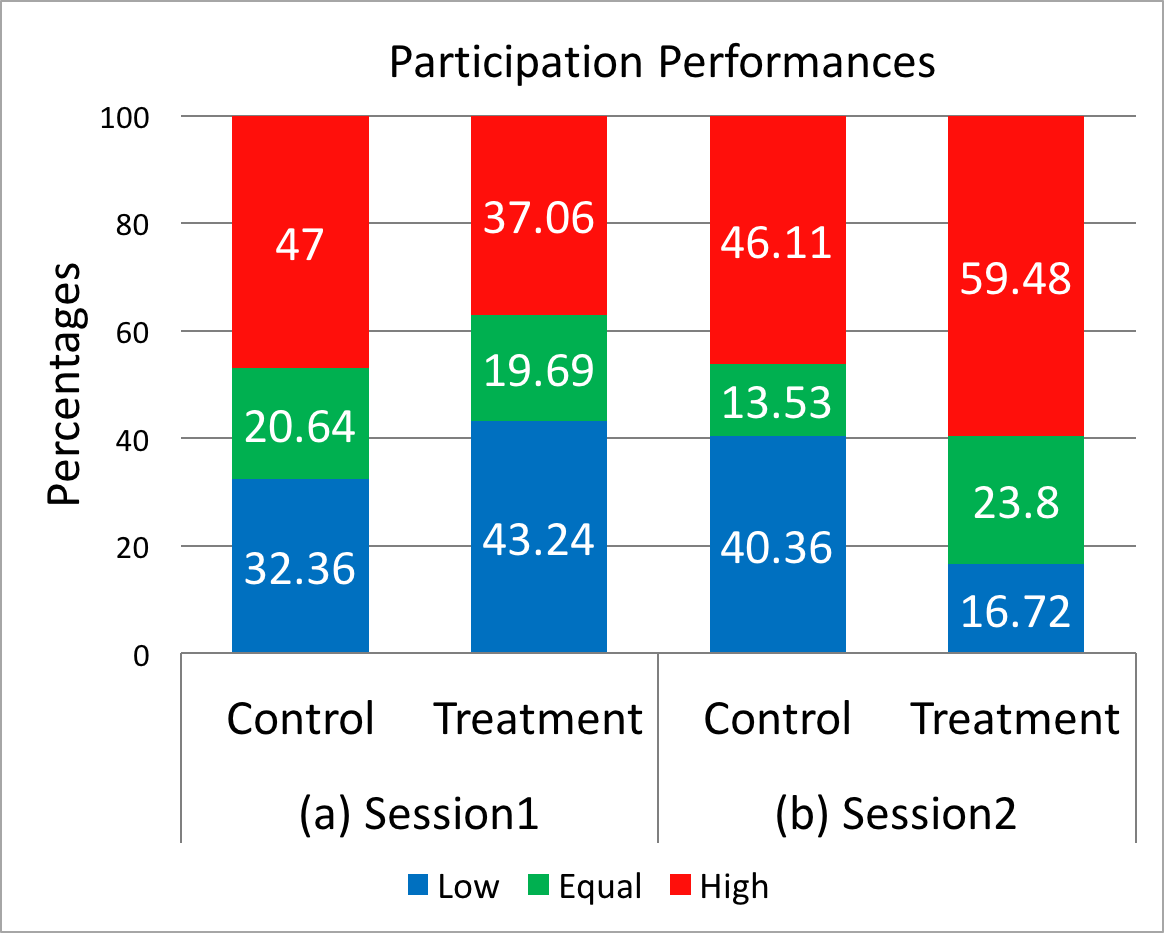}\label{fig:part_perf}}
\hspace{0.05mm}
\subfigure[Facial Emotion Performances]{\includegraphics[width=0.48\columnwidth]{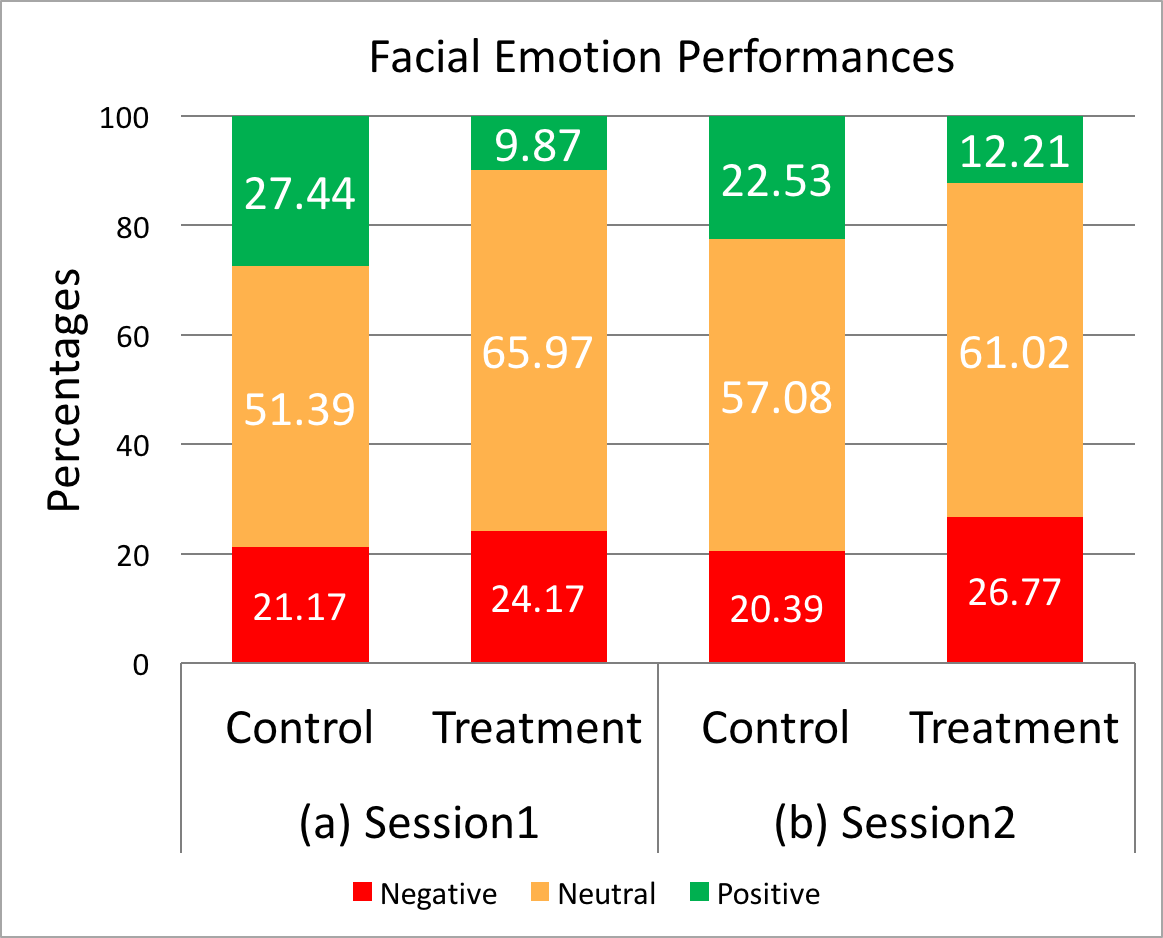}\label{fig:fac_emo_perf}}
\caption{Performance comparison for Control and Treatment Groups in All Sessions}
\vspace{-0.5cm}
\end{figure}


For \textit{low performance} metric, in session-1 the treatment and control groups measures where $(mean_{control_{s1}}=32.36,$ $sd_{control_{s1}}=34.15)$, and $(mean_{treatment_{s1}}=43.24,$ $sd_{treatment_{s1}}=38.98)$, respectively. Fig~\ref{fig:part_perf} shows the relative percentage values of participation feature. The pattern here shows that the treatment group had a tendency to \textit{talk less}, whereas in comparison control group used to \textit{talk more}. The interview with the treatment group participants reveals that, the feedback on \textit{interruption} and  \textit{participation} made them self-conscious. Especially, they did not want the \textit{interruption} counter to go up even during \textit{handovers}, and thus there were more awkward pauses during conversation turnovers.

\begin{quote}
  \textit{D3\_treatment: ``I do feel like I talked a little bit to much in session 2, mainly because I had to really struggle not to talk too much in session 1 and was constantly watching the meter and attempting to get other people to talk so that it wasn't too high for me.''}
\end{quote}

For \textit{positive facial emotion}  metric, in session-1 the treatment group held less positive score ($mean_{treatment_{s1}}=9.86,$ $sd_{treatment_{s1}}=12.87$) than the control group ($mean_{control_{s1}}: 27.44,$ $sd_{control_{s1}}=32.03$) with $F(1,38)=4.937, p=0.033$. In Fig~\ref{fig:fac_emo_perf} we observe the pattern of their facial emotion ranges. It shows that the treatment group remained more in the \textit{neutral emotion} zone and compressed the \textit{positive emotion} zone in comparison with the control group. We investigate the reason for such an effect from the interview responses of the participants. The treatment group participants expressed that the topic being a serious issue they tried to keep the facial emotion score more within the \textit{neutral} zone instead of trying to \textit{smile more}. 

The \textit{interruption} feature, as discussed above, the real-time feedback increased self-conscious among the treatment group participants. Therefore they had difficulty during communication, especially with handovers. Once a speaker stopped talking, the other participants did not have much cue about who would speak next. While trying to grab the floor, participants were worried about overlapping with someone else. This would have the consequence of their \textit{interruption} counters go high.  As a result, there was some hesitancy during handovers, which led to awkward pauses. As stated by a participant:

\begin{quote}

  \textit{B7\_treatment: "In the first session there were a couple of times when there were awkward silence. People were waiting, like, am I interrupting someone, am I talking over someone."}
\end{quote}

\vspace{0.15cm}
As a summary, we find - \textit{Real-time feedback on certain features infuses tendency to talk less, and creates more emotion awareness.}

\subsubsection{\textbf{\hspace{0.1cm}Evaluation of RQ2a}}

\begin{figure}[t]
\centering
\includegraphics[width=0.45\textwidth]{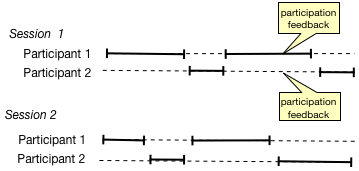}
\caption{Example of Treatment Group Participant Behavior. Participants receiving feedback on a particular feature in session-1 change their behavior-alignment of that feature during session-2 in accordance with the previously received feedback.}\label{fig:stimeline}
\end{figure}

For treatment group, \textit{participation} rate for \textit{low} metric decreased from session-1 ($mean_{treatment_{s1}}=43.24,$ $sd_{treatment_{s1}}=38.98$) to session-2 ($mean_{treatment_{s2}}=16.72,$ $sd_{treatment_{s2}}=21.86$) with $F(1,38)=6.69, p=0.014$. For control group, the metric difference of session-1 ($mean_{control_{s1}}=32.36,$ $sd_{control_{s1}}=35.15$) and session-2 ($mean_{control_{s1}}=40.36,$ $sd_{control_{s2}}=40.29$) is not statistically significant. This implies that, the treatment group participants became significantly more expressive in session-2, as comparison shown in Figure \ref{fig:part_perf}. the ice-breaking effect can be an issue, so we now discuss how this effect is not in action here. The first session requires ice-breaking and may cause low participation rate for different groups. However the fact that the next session significantly impacted only the treatment group and not the control illustrates the after-effect of the real-time feedback. The real-time feedback in session-1 made the participant over-conscious about their performances, which was not present in session-2. Thus these participants became more participatory, allowing them to get into the conversation faster in comparison with the control group. Once the participants experienced the system but in the successive session it was not present, that allowed the participants to be more in control of modifying their behavior on their own instead of replying on the system, which improved their expressiveness. An example is shown in Figure \ref{fig:stimeline}. In session-1, participants-1 talks way more and participant-2 does not speak much. Both receive the \textit{participation} feedback, but for different reasons. Feedback shows to participant-1 about high participation, and participant-2 about low participation. Later on in session-2, even though there is no feedback, participant-1 speaks with more turns and participant-2 speaks more. Thus, both participants modify their behavior in next session according to the feedback received from the previous session.

\vspace{0.15cm}
We summarize that - 
\textit{The following session with no-feedback is influenced by the received real-time feedback of the the previous session.}
\vspace{0.1cm}

\vspace{-0.1cm}
\subsubsection{\textbf{\hspace{0.1cm}Evaluation of RQ2b }}


In session-2 for \textit{low participation} metric, the treatment group ($mean_{treatment_{s2}}=16.72,$ $sd_{treatment_{s2}}=21.86$) remains less in that region in comparison with the control group ($mean_{control_{s2}}=40.36,$ $sd_{control_{s2}}=40.29$) with $F(1,38)=5.054, p=0.03$. It means that the participants in the treatment group were more expressive in the second session than that of the control group participants in session-2. The only influence that differed between the two sessions of the groups was the real-time feedback. In session-2, when both groups had no feedback, the treatment group achieved significant expressiveness than that of the control group. The treatment group participants related the feedback features with respect like this:

\begin{quote}

  \textit{B7\_treatment: "When it comes to respect I think first session went better because there were less interruptions and talking over each other. However, in the second discussion there was much higher engagement from the group and almost no awkward silence while there still being a decent level of respect."}
\end{quote}
\begin{quote}

  \textit{D1\_treatment: "First session (was more respectful). Because interruptions were counted, we tried to let others finish their sentences first."}
\end{quote}

\vspace{0.15cm}
The summarized finding is - 
\textit{The reflective feedback (i.e., propagation effect of the real-time feedback) creates a tendency to be more expressive in comparison with the group receiving no feedback at all.}
\vspace{0.15cm}

\subsection{Self-reported Data Analysis}
After each session, both the control and the treatment group provided the survey responses. Table~\ref{tab:ques} shows the survey questionnaires common for both control and treatment groups. Session-2 contains a set of additional questionnaires which we discuss below as well. To observe RQ1, we compare the average response scores of control and treatment groups after session-1. To observe RQ2a and RQ2b, we compare the average response scores provided by control group in session-1 and session-2, and the same for the treatment group. We apply 2 (control/treatment) X 2 (session-1/session-2) two-way ANOVA with $\alpha=0.05$. As there are 15 questions, to scale the p-value appropriately we apply Bonferroni Correction by multiplying the resultant p-value by 15. After Bonferroni Correction, even though the average responses showed patterns, the significance of the p-value was lost.

\begin{figure}[bh]
\centering
\includegraphics[width=0.3\textwidth]{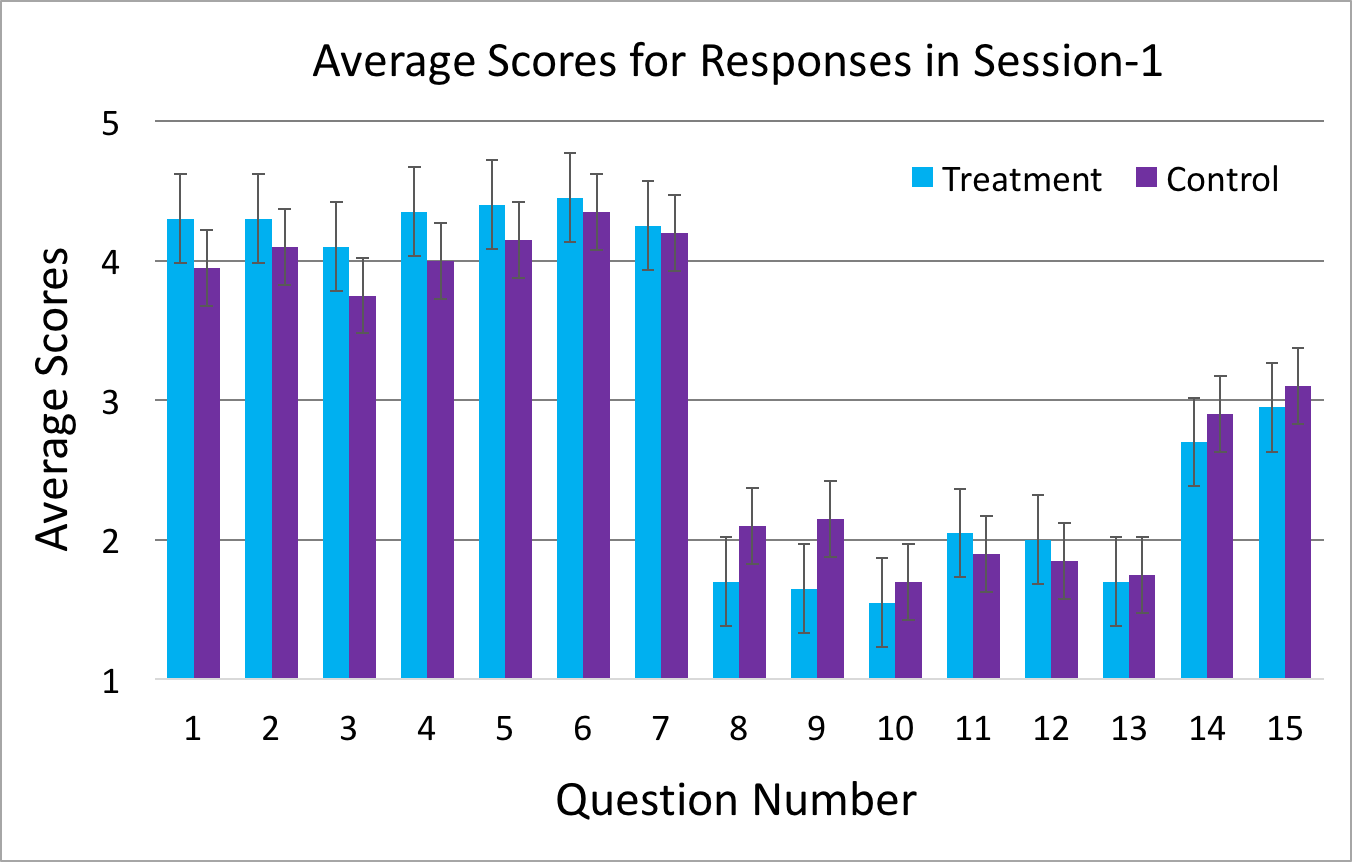}
\caption{Average Scores of Table \ref{tab:ques} Responses in Session-1}\label{fig:avg_t_c_que}
\end{figure}

\begin{figure}[t]
\centering     
\subfigure[Control Group in Session-1,2]{\includegraphics[width=0.5\columnwidth]{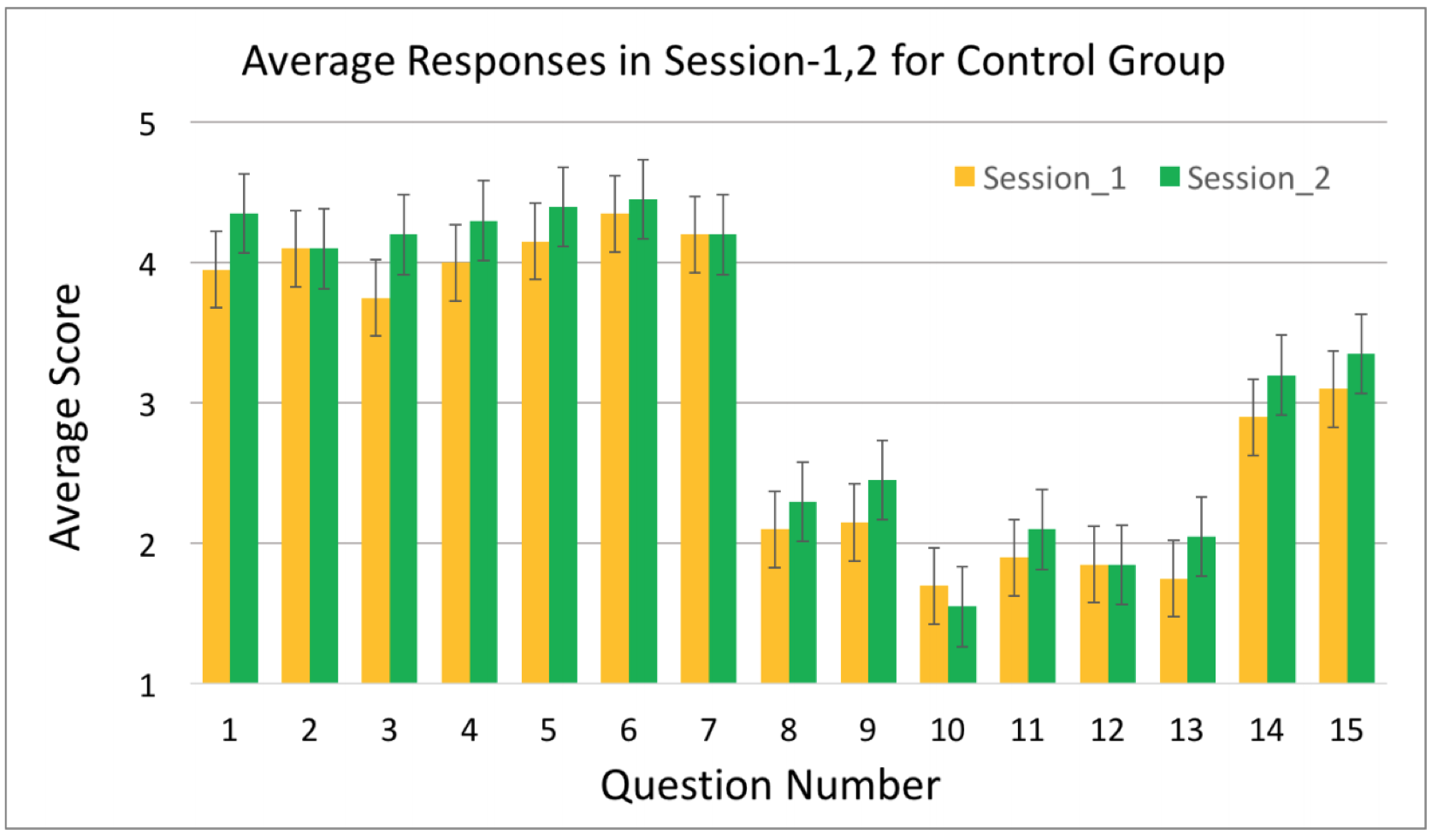}\label{fig:avg_c_s1_s2_que}}
\subfigure[Treatment Group in Session-1,2]{\includegraphics[width=0.47\columnwidth]{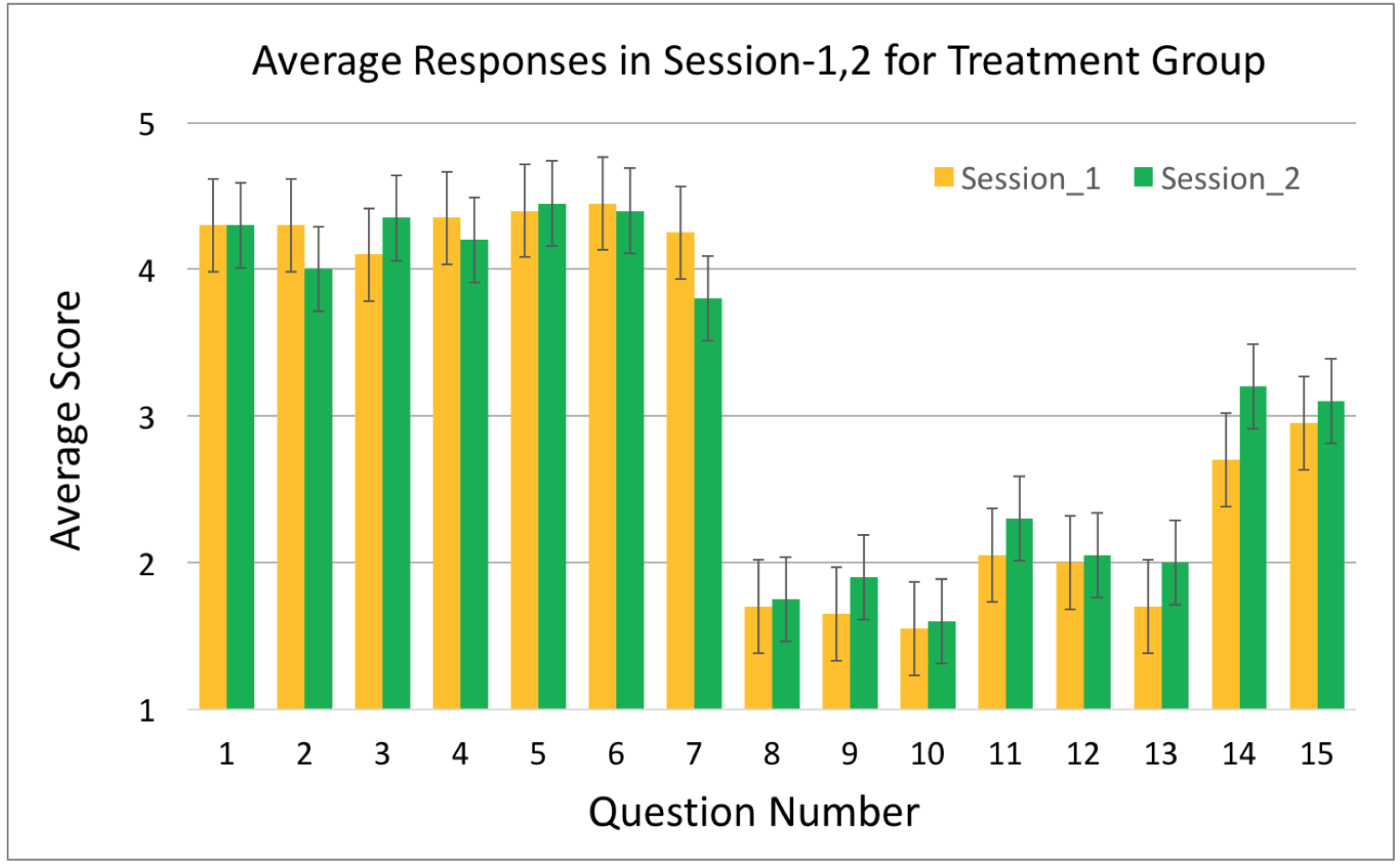}\label{fig:avg_t_s1_s2_que}}
\caption{Average Score Comparison of Table \ref{tab:ques} Responses in Session-1,2}
\end{figure}

For Table~\ref{tab:ques} responses in session-1, the participants of the control and the treatment groups did not report any statistically significant differences. However, on average the treatment group experienced more satisfaction with respect to perceived performance. Fig~\ref{fig:avg_t_c_que} shows the average responses. This is interesting because it shows that our real-time feedback did not instantly affect the responses of the treatment group. Fig~\ref{fig:avg_c_s1_s2_que} shows the comparison of the session-1 and 2 responses by the control group participants.

\begin{table}[bh]
    \caption {Survey Questionnaires} \label{tab:ques} 
\begin{center}
  \begin{tabular}{ l | l }
    \toprule
Q\#	&	Statement\\ \toprule
1 &	I am satisfied with the group discussion.	\\ \hline	
2 &	I was satisfied with the group decision.		\\ \hline
3 &	I am satisfied with how I handled the discussion.		\\ \hline
4 &	I am satisfied with how others handled the \\
 &  discussion.		\\ \hline
5 &	My opinion was heard. \\ \hline		
6 &	My opinion was respected.	\\ \hline	
7 &	The discussion decision found a common ground.	\\ \hline	
8 &	I raised my voice a lot.		\\ \hline
9 &	Others raised their voice a lot.		\\ \hline
10 &	I was unable to express my opinion.		\\ \hline
11	 & Someone else in the group was unable to express \\ 
 & their opinion.	\\ \hline	
12	 &	I interrupted a lot. 	\\ \hline	
13	 &	People interrupted each other a lot.	\\ \hline	
14	 &	I expressed a lot of positive emotion.		\\ \hline
15	& The group expressed a lot of positive emotion.		\\ 
\bottomrule

  \end{tabular}
\end{center}
\vspace{-0.5cm}
\end{table}

On average, the scores imply that in session-2 the control group participants felt better about their performances. When compared the two session responses of the treatment group, the result shows that the treatment group reported to be more positive in session-2. The treatment group participants mentioned that the absence of the real-time feedback made them more spontaneous which they marked as being positive. The discussion topics, in cases, also effected how the participants felt about the session.

\section{Discussion}
\subsection{Findings}

Our observations of the user study reveal several interesting insights on design considerations for collaborative systems. Once crucial issue is setting the behavior thresholds for the features. For example, the level of \textit{high volume} for a loud-spoken person can be different than that of a soft-spoken person. Soft spoken persons, even if they speak on their loudest volume, may not become loud enough. Thus they can get talked over by any other member. It is harder to enforce a generalized threshold for the group. Moreover, modifying the behavior according to a particular setting is not easy either. If a user wants to change a particular behavior, it needs practice sessions with the system to get used to it. In this work we applied a general threshold as the benchmark. We propose that the real-time feedback system should have flexible thresholds as well. This system can be a practice tool for the users to achieve their desired behaviors which they can carry onto other discussions even without the system.

\begin{quote}
  \textit{D10\_control: "I spoke last, which made me feel slightly left out.  I do not like to disagree with people and I am not a strong public speaker, so I did not speak as much as I would have hoped."}
\end{quote}

The real-time feedback, of course, bears trust issues. If one feature is incorrect or inappropriate, it negatively impacts how the users view all the other features. This is particularly hard while identifying appropriate interruption or speech overlap. \textit{Handover} is when the current speaker finishes speaking and the next person takes the floor. It is not considered as an interruption by the people involved. However, it is difficult to determine whether a person is deliberately releasing the speaker's floor or not. Other phenomenons like spontaneous laughter, jinx, etc. are positively related with the conversation. If they get termed as interruptions then the users grow the tendency to avoid the feedback even for valid interruptions. So the system requires better understanding of the social cues and the language uses.

The real-time feedback, especially the \textit{interruption} feature, made the discussion less spontaneous as the participants did not want the \textit{interruption} counters to go high. However, the treatment group participants realized that there were more interruptions in session-2 with no feedback. They saw the interruption as a sign of better engagement. Some of them also agreed that this may be difficult for real life heated debates where interruption may not be a sign of engagement. Thus keeping this feature in the interface creates a trade-off between respectful behavior and engagement. Another case can be that the unfamiliarity with these type of feedback interfaces may be responsible for not understanding how to handle the features. Here are examples of treatment group participants expressing how natural the discussion felt without the feedback -

\begin{quote}

  \textit{D7\_treatment: "In the first session I was looking at the interruption thing, like how many times times I interrupted or something like that. But in second session it wasn't there so it was more of a natural conversation."}
\end{quote}
\begin{quote}

  \textit{C7\_treatment: "I felt like in session-2 we focused more on topic and were thinking about the topic, not on if I am being too loud or being to quiet or interrupting too much. It (the feedback) was useful but not practically necessary."}
\end{quote}

The participants did express that the general agreement or the absence of disagreement gives a vibe of respect. They also expressed that even when the group did not reach an agreement, not receiving "misbehavior" from the group members provided a sense of respect. We propose including feedback on the usages of agreeable or positive words. As mentioned by one participant:

\begin{quote}

  \textit{A10\_control: "I think probably session-1 went more respectfully, possibly just because we agreed so quickly. This session there were some opposing opinions and I think I held the least liked opinion, so though I didn't feel attacked, I don't necessarily agree with the group's decision."}
\end{quote}

Our study finds that real-time feedback  makes the ongoing conversation restrictive and less spontaneous, it reduces undesired behaviors. This is also supported by related literature \cite{tanveer2015, ofek2013, Tausczik} that real-time feedback is generally distracting for in interaction. However, our findings suggest that the real-time feedback has a propagating effect on future conversations. This opens up an opportunity to use real-time feedback as a practice session to prime the users so that the effects can propagate to the actual   discussions. Imagine a self-driven practice session prior to a video call to be more aware of respectful conversational etiquette. Our findings support the assertion that while the real-time feedback is generally distracting, when used appropriately, it would have a positive impact. 

\vspace{-0.1cm}
\subsection{Limitations}

The age range of our study group is 18-24 years. This limited reach does not represent all age ranges. Alongside, forming groups with different aged people in the discussion can generate different dynamics in the discussion. The age, gender, race and their combination are required to understand the general effect of our feedback interface.


To identify the existence of the immediate carryover effect, the sessions were conducted back to back. We highly consider exploring the duration of the effect in the future, as this work was to first verify the existence of the carryover effect. As this study has shown the presence of the carryover effect, the system can now even be a useful tool to practice right before a potentially critical heated discussion.

Besides applying simple visualization techniques, another way our system attempts to control cognitive overload is by controlling the number of features. Prior work emphasizes that the real-time feedback calls for a simplistic design with a limited number of features to impose less cognitive overload. However, the cognitive overload imposed by any real-time system cannot be ignored. In future, we intend to explore how cognitive overload fluctuates with varying number of feedback features and design strategies.

As our objective is to explore the real-time effect of the system, the impact of the visualization property needs to be stabilized across the groups. Therefore, the control group should also be presented with a visual stimulus for receiving the related feature information, except that it would not be in real-time. For this reason, we choose to show a highly influential TEDTalk video to the control group participants before the first discussion session begins. Any video itself is a visual stimulus; on top of that the TEDTalk video also has the speaker showing slides where information was presented using visualization. Thus, by having visual stimuli for both groups, the only observable factor remains to be the real-time nature of stimuli.

Our thresholds are set based on small scale tests, not computed from creating and analyzing a full dataset of group conversations. Moreover, the same threshold may not be appropriate for other types of discussions, for example- tutoring sessions. Even though in our experiment, we kept the hardware setup consistent, the same configuration may not generalize. We plan to deploy the system in the wild to capture a wide range of discussions characteristics.

\vspace{-0.1cm}

\subsection{Future Work}

Even though we designed the interface for debate-like discussions, its effects can be observed for other types of conversations as well. This interface can help include or exclude features depending on the discussion type to observe whether any common properties are prevalent for other discussions with the same real-time feedback system.

With any system providing automated real-time feedback, there is a need for transparency and gaining user's trust. For example, some users, by default, may believe that the system has made an error, or the participants' may just want to understand the reason for receiving a certain feedback. To address this, our future work will involve individual post-feedback for the participants by summarizing the frequency and the context in which the feedback was generated. The users will also be able to indicate whether a piece of feedback was effective or not, allowing the system to continuously improve further on.


Appropriate interpretations of the nonverbal feedback remains an active area of exploration. For example, while interruptions may appear disruptive among strangers, but for close friends, interruptions are often expected and they are not detrimental to the conversation. How a system would recognize and interpret such interruptions remain an open problem. In our system, our feedback design had an implicit bias towards equal participation. However, in many conversations, a natural leader with more expertise can emerge as a leader adding important perspective to the conversations. Our future work will involve automated adaptation of such emergence.    

In our experiment, each of the discussion sessions were 15-minute long. Some participants stated that sometimes it was not enough for a deeper conversation. Extending the time period may add more characteristics to the conversations. Experimenting with longer sessions remains part of our future work. Our public codebase will also help the affective computing community analyzing longer discussion sessions among people having different rapport.

Cognitive overload is a crucial property to measure, especially for conversational settings such as tutoring (e.g., MOOCs), telehealth (e.g., remote doctor-patient conversation), etc. By modifying our public platform for measuring cognitive load from videos, the interested researchers in the related fields can not only assess performance during an ongoing conversation, but also modify and apply the appropriate real-time feedback for the speakers involved.

Another interesting aspect regarding cognitive overload would be to modify the number and the design of the feedback feature to observe the effect on cognitive overload. To measure the level of overload, either the platform can be modified for automated sensing of cognitive overload from video, or user provided self-reported data on cognitive load scale can be used. Thus, the system bears the potential to provide deeper insights on the real-time feedback and cognitive overload.

\vspace{-0.08cm}
\section{Conclusion}
In this paper, we present our video conferencing system incorporated with real-time feedback. We observe the instant and long-term reflective effect of the real-time feedback within a debate-like group discussion. We conduct a treatment-control user study with 40 participants in a total of 20 discussion sessions to compare the effects. Our results show how real-time feedback reduces spontaneity of the discussion for the video conferencing platform, but influences the expressiveness in the following discussion without any feedback. The implications can be useful for research using real-time feedback and videoconferencing based group discussions.
Due to COVID-19, all of our interactions are taking place online. It has further highlighting the importance of understanding the nonverbal nuances and conversational dynamics in videocalls. Some of the fundamental aspects of building relationships like establishing rapport, showing empathy, sincerely listening to each other do not translate effectively. It is easy to misread cues resulting in unpleasant exchanges. There is an opportunity to design interventions to help individuals cope with new normal way of communicating online. This paper is an initial exploration towards that direction.

\section*{Acknowledgment}
This work was supported by the National Science Foundation Award IIS-1464162, a Google Faculty Research Award, and Microsoft Azure for Research grant.

\bibliographystyle{IEEEtran}
\bibliography{IEEEabrv,realtimecocoTAC20}

\begin{thebibliography}{10}
\providecommand{\url}[1]{#1}
\csname url@samestyle\endcsname
\providecommand{\newblock}{\relax}
\providecommand{\bibinfo}[2]{#2}
\providecommand{\BIBentrySTDinterwordspacing}{\spaceskip=0pt\relax}
\providecommand{\BIBentryALTinterwordstretchfactor}{4}
\providecommand{\BIBentryALTinterwordspacing}{\spaceskip=\fontdimen2\font plus
\BIBentryALTinterwordstretchfactor\fontdimen3\font minus
  \fontdimen4\font\relax}
\providecommand{\BIBforeignlanguage}[2]{{%
\expandafter\ifx\csname l@#1\endcsname\relax
\typeout{** WARNING: IEEEtran.bst: No hyphenation pattern has been}%
\typeout{** loaded for the language `#1'. Using the pattern for}%
\typeout{** the default language instead.}%
\else
\language=\csname l@#1\endcsname
\fi
#2}}
\providecommand{\BIBdecl}{\relax}
\BIBdecl

\bibitem{Mansbridge2009}
J.~Mansbridge, J.~Bohman, S.~Chambers, D.~Estlund, A.~Follesdal, A.~Fung,
  C.~Lafont, B.~Manin, and J.~l. Marti, ``The place of self-interest and the
  role of power in deliberative democracy,'' \emph{Journal of Political
  Philosophy}, vol.~18, no.~1, 2009.

\bibitem{haidt2012}
J.~Haidt, \emph{The Righteous Mind: Why Good People Are Divided By Politics and
  Religion}.\hskip 1em plus 0.5em minus 0.4em\relax New York: Pantheon Books,
  2012.

\bibitem{derby2006agile}
E.~Derby, D.~Larsen, and K.~Schwaber, \emph{Agile retrospectives: Making good
  teams great}.\hskip 1em plus 0.5em minus 0.4em\relax Pragmatic Bookshelf,
  2006.

\bibitem{sellen1995}
A.~J. Sellen, ``Remote conversations: The effects of mediating talk with
  technology,'' \emph{Human-Computer Interaction (CHI)}, pp. 401--444, 1995.

\bibitem{tanveer2015}
M.~I. Tanveer, E.~Lin, and M.~E. Hoque, ``{Rhema: A Real-time in-situ
  Intelligent Interface to Help People with Public Speaking},'' in
  \emph{Proceedings of the 20th International Conference on Intelligent User
  Interfaces (IUI '15)}, 2015, pp. 286--295.

\bibitem{ofek2013}
E.~Ofek, S.~T. Iqbal, and K.~Strauss, ``{Reducing Disruption from Subtle
  Information Delivery During a Conversation: Mode and Bandwidth
  Investigation},'' in \emph{{Proceedings of the SIGCHI Conference on Human
  Factors in Computing Systems (CHI '13)}}, 2013, pp. 3111--3120.

\bibitem{homer2008}
B.~D. Homer, J.~L. Plass, and L.~Blake, ``{The Effects of Video on Cognitive
  Load and Social Presence in Multimedia-Learning},'' \emph{{Computers in Human
  Behavior}}, vol.~24, no.~3, pp. 786--797, 2008.

\bibitem{maia2016}
R.~C. Maia and T.~A.~S. Rezende, ``Respect and disrespect in deliberation
  across the networked media environment: Examining multiple paths of political
  talk,'' \emph{Journal of Computer-Mediated Communication}, vol.~21, no.~2,
  2016.

\bibitem{James1992}
J.~E. Driskell and E.~Salas, ``Collective behavior and team performance,''
  \emph{Human Factors}, vol.~34, no.~3, 1992.

\bibitem{dunbar2005}
N.~E. Dunbar and J.~K. Burgoon, ``Perceptions of power and interactional
  dominance in interpersonal relationships,'' \emph{Journal of Social and
  Personal Relationships}, 2005.

\bibitem{lamb1981nonverbal}
T.~A. Lamb, ``Nonverbal and paraverbal control in dyads and triads: Sex or
  power differences?'' \emph{Social Psychology Quarterly}, pp. 49--53, 1981.

\bibitem{burgeon2002nonverbal}
J.~Burgeon and G.~Hoobler, ``Nonverbal signals. handbook of interpersonal
  communication 3rd. edition,'' 2002.

\bibitem{dunbar2008}
N.~Dunbar, A.~M.~Bippus, and S.~Young, ``Interpersonal dominance in relational
  conflict: A view from dyadic power theory,'' \emph{Interpersona : An
  International Journal on Personal Relationships}, vol.~2, 10 2008.

\bibitem{Malte2016}
M.~F. Jung, ``Coupling interactions and performance: Predicting team
  performance from thin slices of conflict,'' \emph{ACM Trans. Comput.-Hum.
  Interact.}, vol.~23, no.~3, Jun. 2016.

\bibitem{anderson1999}
L.~Andersson and C.~M.~Pearson, ``Tit for tat? the spiraling effect of
  incivility in the workplace,'' \emph{The Academy of Management Review},
  vol.~24, pp. 452--471, 07 1999.

\bibitem{picard2001}
R.~W. {Picard}, E.~{Vyzas}, and J.~{Healey}, ``Toward machine emotional
  intelligence: analysis of affective physiological state,'' \emph{IEEE
  Transactions on Pattern Analysis and Machine Intelligence}, vol.~23, no.~10,
  pp. 1175--1191, Oct 2001.

\bibitem{rothenberg1971anger}
A.~Rothenberg, ``On anger,'' \emph{American Journal of Psychiatry}, vol. 128,
  no.~4, pp. 454--460, 1971.

\bibitem{sullivan1973clinical}
H.~S. Sullivan, \emph{Clinical studies in psychiatry}.\hskip 1em plus 0.5em
  minus 0.4em\relax WW Norton \& Company, 1973, vol.~2.

\bibitem{Costa2018}
J.~Costa, M.~F. Jung, M.~Czerwinski, F.~Guimbreti\`{e}re, T.~Le, and
  T.~Choudhury, ``Regulating feelings during interpersonal conflicts by
  changing voice self-perception,'' in \emph{Proceedings of the 2018 CHI
  Conference on Human Factors in Computing Systems}, 2018.

\bibitem{ab01}
P.~L. McLeod, J.~K. Liker, and S.~A. Lobel, ``{Process Feedback in Task Groups:
  An Application of Goal Setting},'' \emph{The Journal of Applied Behavioral
  Science}, vol.~28, no.~1, 1992.

\bibitem{Neubert1998}
M.~R.~Barrick, G.~L.~Stewart, M.~Neubert, and M.~Mount, ``Relating member
  ability and personality to work-team processes and team effectiveness,''
  \emph{Journal of Applied Psychology}, 1998.

\bibitem{Nadler2003}
J.~Nadler, ``{Rapport in negotiation and conflict resolution},''
  \emph{{Marquette Law Review}}, vol. 285, no. 1990, 2003.

\bibitem{Kihara2016}
H.~Kihara, S.~Fukushima, and T.~Naemura, ``Analysis of human nodding behavior
  during group work for designing nodding robots,'' in \emph{Proceedings of the
  19th International Conference on Supporting Group Work}, ser. GROUP '16,
  2016.

\bibitem{david2011}
D.~A. Huffaker, R.~Swaab, and D.~Diermeier, ``The language of coalition
  formation in online multiparty negotiations,'' \emph{Journal of Language and
  Social Psychology}, vol.~30, no.~1, 2011.

\bibitem{rawassizadeh2015}
R.~Rawassizadeh, M.~Tomitsch, M.~Nourizadeh, E.~Momeni, A.~Peery, L.~Ulanova,
  and M.~Pazzani, ``{Energy-Efficient Integration of Continuous Context Sensing
  and Prediction into Smartwatches},'' \emph{Sensors}, vol.~15, no.~9, pp.
  22\,616--22\,645, 2015.

\bibitem{zeng2009}
Z.~Zeng, M.~Pantic, G.~I. Roisman, and T.~S. Huang, ``{A Survey of Affect
  Recognition Methods: Audio, Visual, and Spontaneous Expressions},''
  \emph{{IEEE Transactions on Pattern Analysis and Machine Intelligence}},
  vol.~31, no.~1, pp. 39--58, 2009.

\bibitem{bailenson2007}
{Bailenson, Jeremy N and Yee, Nick and Brave, Scott and Merget, Dan and Koslow,
  David}, ``{Virtual interpersonal touch: expressing and recognizing emotions
  through haptic devices},'' \emph{{Human-Computer Interaction}}, vol.~22,
  no.~3, pp. 325--353, 2007.

\bibitem{keltner2003facial}
D.~Keltner, P.~Ekman, G.~C. Gonzaga, and J.~Beer, ``Facial expression of
  emotion.'' 2003.

\bibitem{Leavitt1951}
H.~J. Leavitt, ``{Some Effects of Certain Communication Patterns on Group
  Performance},'' \emph{{Journal of Abnormal Psychology}}, 1951.

\bibitem{Diamant2009}
E.~I. Diamant, B.~Y. Lim, A.~Echenique, G.~Leshed, and S.~R. Fussell,
  ``Supporting intercultural collaboration with dynamic feedback systems:
  Preliminary evidence from a creative design task,'' in \emph{Extended
  Abstracts on Human Factors in Computing Systems}, ser. CHI EA '09, 2009.

\bibitem{Burgoon1991}
J.~Burgoon and D.~Newton, ``\BIBforeignlanguage{English (US)}{Applying a social
  meaning model to relational message interpretations of conversational
  involvement: Comparing observer and participant perspectives},''
  \emph{\BIBforeignlanguage{English (US)}{The Southern Communication Journal}},
  vol.~56, no.~2, pp. 96--113, 1991.

\bibitem{Boyd2016}
L.~Boyd, A.~Rangel, H.~Tomimbang, A.~Conejo-Toledo, K.~Patel, M.~Tentori, and
  G.~Hayes, ``{SayWAT: Augmenting Face-to-Face Conversations for Adults with
  Autism},'' in \emph{{Proceedings of the 2016 CHI Conference on Human Factors
  in Computing Systems (CHI '16)}}, 2016, pp. 4872--4883.

\bibitem{hegde2015}
N.~Hegde, G.~D. Fulk, and E.~S. Sazonov, ``{Development of the RT-GAIT, a
  Real-Time Feedback Device to Improve Gait of Individuals with Stroke},'' in
  \emph{{37th Annual International Conference of the IEEE Engineering in
  Medicine and Biology Society (EMBC '15)}}, 2015, pp. 5724--5727.

\bibitem{rawassizadeh2018}
R.~Rawassizadeh, T.~J. Pierson, R.~Peterson, and D.~Kotz, ``{NoCloud: Exploring
  Network Disconnection through On-Device Data Analysis},'' \emph{{IEEE
  Pervasive Computing}}, vol.~17, no.~1, pp. 64--74, 2018.

\bibitem{RN11}
M.~Tentori and G.~R. Hayes, ``{Designing for Interaction Immediacy to Enhance
  Social Skills of Children with Autism},'' in \emph{{Proceedings of the 12th
  ACM international conference on Ubiquitous Computing (Ubicomp '12)}}, 2012,
  pp. 51--60.

\bibitem{campbell_}
S.~W. Campbell and N.~Kwak, ``Mobile communication and civil society: Linking
  patterns and places of use to engagement with others in public,'' \emph{Human
  Communication Research}, vol.~37, 2011.

\bibitem{Tausczik}
Y.~R. Tausczik and J.~W. Pennebaker, ``Improving teamwork using real-time
  language feedback,'' in \emph{Proceedings of the SIGCHI Conference on Human
  Factors in Computing Systems (CHI '13)}, 2013, pp. 459--468.

\bibitem{grimmett1988reflection}
P.~P. Grimmett and G.~Erickson, ``Reflection in teacher education,'' 1988.

\bibitem{husu2007developing}
J.~Husu, S.~Patrikainen, and A.~Toom, ``Developing teachers’ competencies in
  reflecting on teaching,'' in \emph{Making a Difference}.\hskip 1em plus 0.5em
  minus 0.4em\relax Brill Sense, 2007, pp. 125--140.

\bibitem{leshed2013}
N.~Goyal, G.~Leshed, and S.~R. Fussell, ``{Leveraging Partner's Insights for
  Distributed Collaborative Sensemaking},'' in \emph{Proceedings of the 2013
  conference on Computer supported cooperative work companion (CSCW '13)},
  2013.

\bibitem{Leshed2009}
E.~I. Diamant, B.~Y. Lim, A.~Echenique, G.~Leshed, and S.~R. Fussell,
  ``{Supporting Intercultural Collaboration with Dynamic Feedback Systems:
  Preliminary Evidence from a Creative Design Task},'' in \emph{CHI '09
  Extended Abstracts on Human Factors in Computing Systems}.

\bibitem{Nowak2012}
M.~Nowak, J.~Kim, N.~W. Kim, and C.~Nass, ``Social visualization and
  negotiation: Effects of feedback configuration and status,'' in
  \emph{Proceedings of the ACM 2012 Conference on Computer Supported
  Cooperative Work}, ser. CSCW '12, 2012.

\bibitem{leshed2007}
G.~Leshed, J.~T. Hancock, D.~Cosley, P.~L. McLeod, and G.~Gay, ``{Feedback for
  Guiding Reflection on Teamwork Practices},'' in \emph{{Proceedings of the
  2007 international ACM conference on Supporting group work (GROUP '07)}},
  2007.

\bibitem{sellen2010}
D.~S. Kirk, A.~Sellen, and X.~Cao, ``{Home video communication: mediating
  'closeness'},'' in \emph{Proceedings of the 2010 ACM Conference on Computer
  Supported Cooperative Work (CSCW '10)}, 2010, pp. 135--144.

\bibitem{buhler2013and}
T.~Buhler, C.~Neustaedter, and S.~Hillman, ``How and why teenagers use video
  chat,'' in \emph{Proceedings of the 2013 conference on Computer supported
  cooperative work}.\hskip 1em plus 0.5em minus 0.4em\relax ACM, 2013, pp.
  759--768.

\bibitem{ames2010making}
M.~G. Ames, J.~Go, J.~Kaye, and M.~Spasojevic, ``Making love in the network
  closet: the benefits and work of family videochat,'' in \emph{Proceedings of
  the 2010 ACM conference on Computer supported cooperative work}.\hskip 1em
  plus 0.5em minus 0.4em\relax ACM, 2010, pp. 145--154.

\bibitem{judge2010sharing}
T.~K. Judge and C.~Neustaedter, ``Sharing conversation and sharing life: video
  conferencing in the home,'' in \emph{Proceedings of the SIGCHI Conference on
  Human Factors in Computing Systems}.\hskip 1em plus 0.5em minus 0.4em\relax
  ACM, 2010, pp. 655--658.

\bibitem{Losada1990}
M.~Losada, P.~Sanchez, and E.~E. Noble, ``Collaborative technology and group
  process feedback: Their impact on interactive sequences in meetings,'' in
  \emph{Proceedings of the 1990 ACM Conference on Computer-supported
  Cooperative Work}, ser. CSCW '90, 1990.

\bibitem{mcgrath1994groups}
J.~E. McGrath and A.~B. Hollingshead, \emph{Groups interacting with technology:
  Ideas, evidence, issues, and an agenda.}\hskip 1em plus 0.5em minus
  0.4em\relax Sage Publications, Inc, 1994.

\bibitem{straus1996getting}
S.~G. Straus, ``Getting a clue: The effects of communication media and
  information distribution on participation and performance in
  computer-mediated and face-to-face groups,'' \emph{Small group research},
  vol.~27, no.~1, pp. 115--142, 1996.

\bibitem{Kim2008}
T.~Kim, A.~Chang, L.~Holland, and A.~S. Pentland, ``Meeting mediator: Enhancing
  group collaborationusing sociometric feedback,'' in \emph{Proceedings of the
  2008 ACM Conference on Computer Supported Cooperative Work}, ser. CSCW '08,
  2008.

\bibitem{Kim2012}
T.~Kim, P.~Hinds, and A.~Pentland, ``{Awareness As an Antidote to Distance:
  Making Distributed Groups Cooperative and Consistent},'' in \emph{Proceedings
  of the ACM 2012 Conference on Computer Supported Cooperative Work (CSCW
  '12)}, 2012, pp. 1237--1246.

\bibitem{byun2011}
B.~{Byun}, A.~{Awasthi}, P.~A. {Chou}, A.~{Kapoor}, B.~{Lee}, and
  M.~{Czerwinski}, ``Honest signals in video conferencing,'' in \emph{2011 IEEE
  International Conference on Multimedia and Expo}, 2011.

\bibitem{Faucett2017}
H.~A. Faucett, M.~L. Lee, and S.~Carter, ``I should listen more: Real-time
  sensing and feedback of non-verbal communication in video telehealth,''
  \emph{Proc. ACM Hum.-Comput. Interact.}, vol.~1, no. CSCW, pp. 44:1--44:19,
  Dec. 2017.

\bibitem{Helen2017}
H.~A. He, N.~Yamashita, A.~Hautasaari, X.~Cao, and E.~M. Huang, ``Why did they
  do that?: Exploring attribution mismatches between native and non-native
  speakers using videoconferencing,'' in \emph{Proceedings of the 2017 ACM
  Conference on Computer Supported Cooperative Work and Social Computing}, ser.
  CSCW '17, 2017.

\bibitem{Samrose2018}
S.~Samrose, R.~Zhao, J.~White, V.~Li, L.~Nova, Y.~Lu, M.~R. Ali, and M.~E.
  Hoque, \emph{Proc. ACM Interact. Mob. Wearable Ubiquitous Technol.}, vol.~1,
  no.~4, 2018.

\bibitem{engstrom2017}
J.~Engstr{\''o}m, G.~Markkula, T.~Victor, and N.~Merat, ``{Effects of Cognitive
  Load on Driving Performance: The Cognitive Control Hypothesis},'' \emph{Human
  {F}actors}, vol.~59, no.~5, pp. 734--764, 2017.

\bibitem{tillman2017}
G.~Tillman, D.~Strayer, A.~Eidels, and A.~Heathcote, ``{Modeling Cognitive Load
  Effects of Conversation Between a Passenger and Driver},'' \emph{{Attention,
  Perception, \& Psychophysics}}, vol.~79, no.~6, pp. 1795--1803, 2017.

\bibitem{McDuff2016}
D.~McDuff, A.~Mahmoud, M.~Mavadati, M.~Amr, J.~Turcot, and R.~e. Kaliouby,
  ``Affdex sdk: A cross-platform real-time multi-face expression recognition
  toolkit,'' in \emph{Proceedings of the 2016 CHI Conference Extended Abstracts
  on Human Factors in Computing Systems}, ser. CHI EA '16.\hskip 1em plus 0.5em
  minus 0.4em\relax New York, NY, USA: ACM, 2016.

\end{thebibliography}

\clearpage

\end{document}